\documentstyle[a4]{article}
\begin{document}

\vspace{0.3cm}
\begin{center}
{\Large {\bf Note on Triangle Anomalies and Assignment of Singlet\\
 in 331-like Model}}
\end{center}
\vspace{0.8cm}

\begin{center}
T. Kiyan\footnote{E-mail : {kyan@sci.kumamoto-u.ac.jp}}, T. Maekawa and S. Yokoi\\
Department of Physics, Kumamoto University, Kumamoto 860-8555 Japan
\end{center}
\begin{abstract}
\noindent 
It is pointed out that in the $331-$like model which uses both fundamental 
and complex conjugate representations for an assignment of the representations 
to the left-handed quarks and the scalar representation to their corresponding 
right-handed counterparts, the nature of the scalar should be taken into account 
in order to make the fermion triangle anomalies in the theory anomaly-free, 
i.e. renormalizable in a sense with no anomalies, 
even after the spontaneous symmetry breaking.
\end{abstract}
\vspace{0.8cm}

In this note, the 331 model\cite{MIN,HL,RHN}, in which the fundamental 
representation for the left-handed quarks in some families and complex conjugate 
representation for those in the other families are assigned together with the singlet 
representation for the corresponding right-handed counterparts, is discussed as 
an example though our discussion can be applied to any such model 
and the primary model\cite{SU} assigning only the fundamental representation to 
the left-handed fields is not treated here. The original Lagrangian of the 331 model 
is constructed to be anomaly-free, in the sense that total charge on 
all particles over three families vanishes, as well as gauge-invariant 
and thus it is expected that the results will be renormalizable. 
However, the renormalizability of the theory should hold not only for 
the original Lagrangian\cite{DP} but also for the Lagrangian written in 
terms of the mass eigenstate (physical) bases after the spontaneous symmetry 
breaking\cite{J,PL}. If the anomaly coefficients for all possible triangle diagrams 
after the symmetry breaking can not be expressed in terms of the trace of 
the product of the representation matrices corresponding to those before 
the symmetry breaking, the anomalies associated with the triangle diagrams 
will remain without vanishing and then the renormalizability of the theory will be 
lost even though the starting Lagrangian is not so\cite{J}. In what follows, 
it is shown that the result of the theory depends through the Yukawa interactions 
on whether the nature of the singlet is taken into account or not \cite{KKM}.

The 331 model\cite{MIN,HL,RHN} is studied by many authors since Pisano 
and Pleitez and many discussions on the lepton sector  approach are made 
phenomenologically from various points of view\cite{NMASS,LPT}. 
We have an interest in the anomalies which are singularities associated with 
the fermion triangle contributions to the vertex of three currents arisen from 
the fermion covariant kinetic energy terms and have no relation to the mass of 
the fermions. Thus, the following assignment of the representations to 
the basic particles in the three families is adopted in the case of 
the $\rm SU(3)_C\otimes SU(3)_L\otimes U(1)_N$ model as follows\cite{HL}
\begin{eqnarray}
&& l^0_{aL}=\left( 
\begin{array}{c} \nu^0_a \\ e^0_a \\  E^0_a 
\end{array} \right)_L
\sim  (1,\ 3, \ 0), \nonumber\\
&& \nu^0_{aR} \sim (1,\ 1,\ 0),\ e^0_{aR} 
\sim  (1,\ 1,\ -1),\ E^0_{aR} 
\sim  (1,\ 1,\ 1),\nonumber\\
&& Q^0_{iL}=\left( 
\begin{array}{c} d^0_i \\ u^0_i \\ J^0_i 
\end{array} \right)_L
\sim  (3,\bar 3, -1/3), \label{1}\\
&& u^0_{iR}\sim (3,\ singlet,\ 2/3),\ d^0_{iR}
\sim (3,\ singlet,\ -1/3),\ J^0_{iR}
\sim (3,\ singlet,\ -4/3),\nonumber\\    
&& Q^0_{3L}=\left( 
\begin{array}{c} u^0_3 \\ d^0_3 \\  J^0_3 
\end{array} \right)_L
\sim  (3,\ 3, \ 2/3), \nonumber\\
&& u^0_{3R}\sim (3,\ singlet,\ 2/3),\ d^0_{3R}
\sim (3,\ singlet,\ -1/3),\ J^0_{3R}
\sim (3,\ singlet,\ 5/3),\nonumber
\end{eqnarray}
where the suffices $a (=1,2,3)$ and $i (=1,2)$ denote the family numbers.
 
And the word $``singlet"$ for the $(u^0_i, d^0_i, J^0_i)_R$ and 
$(u^0_3, d^0_3, J^0_3)_R$ quarks  are used instead of usual ``1", 
though $``1"$ is used in the leptons without any  subscript. This is because in 
the case of leptons only the fundamental representation is assigned to 
the left-handed fields and thus it is obvious that the singlet of 
the right-handed field accompanied with the left-handed one is a scalar 
with respect to the transformation of the fundamental representation. 
On the other hand, the fundamental and complex conjugate representations 
are used for the assignment of the left-handed quarks and then there are 
some possibilities about the singletness (scalar) in physics but not 
mathematics,\ i.e., bosonic, fermionic and antifermionic singlets such as 
the colorless singlets in the quark model. As is well known, 
the singlet appears in the representation theory of $\rm SU(3)$ as follows
\begin{eqnarray}
&& 3\otimes {\bar 3} 
=1\oplus 8 ,({\rm i})\nonumber\\
&& 3\otimes 3 \otimes 3 
=1\oplus 8\oplus 8\oplus 10 ,({\rm ii})\label{2}\\
&& {\bar 3}\otimes {\bar 3} \otimes {\bar 3}
=1\oplus 8\oplus 8\oplus {\overline {10}} .({\rm iii})\nonumber
\end{eqnarray}
The singlets ``1" on the right side are equivalent mathematically or from 
the point view of the transformation of $SU(3)$ with each other due to 
the relation $3\otimes 3= {\bar 3}+6$ but physically should be considered nonequivalent 
because their configurations are different from each other in the form as it stands 
and if only the fundamental representation $3$ is assigned to the basic particles such as 
in the color quark model, the ``1" in (i) $\sim $ (iii) will represent different scalars 
physically\cite{PL,KKM}, e.g., will be called ``bosonic" singlet ($``1"$, meson ) for (i), 
``fermionic" singlet ($``1_3"$, baryon ) for (ii) and ``antifermionic" singlet 
($``1_{\bar 3}"$, antibaryon ) for (iii). If the configurations for these scalars 
as it is are adopted, it may be considered that the bosonic singlet in (i) is the scalar 
with respect to the transformation of $3$ and/or $\bar 3$, the fermionic singlet in (ii) 
is the scalar with respect to the transformation of only $3$ but not $\bar 3$, 
and the antifermionic singlet in (iii) is the scalar with respect to the transformation 
of only $\bar 3$ but not $3$. There is no problem when only the fundamental 
( or complex conjugate ) representation is assigned to the three left-handed quarks as 
in the case of the leptons\cite{SU,PL}. The problem is that though the ``singlet" in 
({\ref 1}) invariant under the transformation of $\rm SU(3)_L$ should not be considered as 
the composite as in ({\ref 2}) and the assignment of the singlet will be free a priori, 
the singlets accompanied with the $3$ and $\bar 3$ representations may be considered 
unique or non-unique. Usually, only the bosonic singlet  $``1"$ is adopted as far as 
we know\cite{MIN,HL,RHN}. It is important for us to distinguish the possibilities 
because the assignment $1_3$ to the right-handed quarks accompanied with 
the $3$ representation and $1_{\bar 3}$ to those with the $\bar 3$ representation 
bring about the reasonable results than the assignment $``1"$ of the bosonic singlet 
to all right-handed quarks as shown below and an adoption of such a transformation 
for the right-handed singlet counterparts seems reasonable from a physical point of view 
in the chiral theory with left-handed fundamental and complex conjugate representations. 

The three scalar fields, $\chi$, $\rho$ and $\eta$, are introduced to break 
the symmetry spontaneously and then give the mass to the fields as follows \cite{HL}
\begin{eqnarray}
\chi &=& \left(
 \begin{array}{c}
 \chi^{-}\\ \chi^{--}\\ \chi^0
 \end{array} \right) \sim (1, 3, -1),\nonumber\\
\rho &=& \left(
 \begin{array}{c}
 \rho^{+}\\ \rho^{0}\\ \rho^{++}
 \end{array} \right)\sim (1,3,1),\label{3}\\
\eta &=& \left(
 \begin{array}{c}
 \eta^{0}\\ \eta^{-}\\ \eta^{+}
 \end{array} \right)\sim (1,3,0).\nonumber
\end{eqnarray}
The charge operator $Q$ is defined by $Q=T_3-{\sqrt 3} T_8 +N$ with 
the generators $T_3, T_8 $  of $\rm SU(3)_L$ together with $T_i$ 
and $N$ of $\rm U(1)_N$. In what follows, the color symmetry is omitted 
because it has no direct relation to our discussion. The invariant Lagrangians for 
the fermions and the Yukawa interactions are given by \cite{MIN,HL,RHN}
\begin{eqnarray}
&& L_f ={\bar l^0}_{aL}i{\ooalign{\hfil/\hfil\crcr$D$}}l^0_{aL}
+({\bar \nu^0}_a,{\bar e^0}_a,{\bar E^0}_a)_R 
i{\ooalign{\hfil/\hfil\crcr$D$}} 
\left( 
\begin{array}{c} \nu^0_a \\ e^0_a \\ E^0_a  \end{array} \right)_R 
+{\bar Q^0}_{aL}i{\ooalign{\hfil/\hfil\crcr$D$}}Q^0_{aL}
+(\bar u^0_a,\bar d^0_a, \bar J^0_a)_R i
{\ooalign{\hfil/\hfil\crcr$D$}} 
\left( 
\begin{array}{c} u^0_a \\  d^0_a \\ J^0_a \end{array} \right)_R ,
\nonumber\\
&& L^\chi_Y =(\bar l^0_1,\bar l^0_2,\bar l^0_3)_L\mu^\chi 
\left( \begin{array}{c} E^0_1 \\  E^0_2 \\ E^0_3 
\end{array} \right)_R\chi 
+\Gamma^\chi_3{\bar Q^0}_{3L}J^0_{3R}\chi \nonumber\\
&& +(\bar Q^0_1,\bar Q^0_2)_L
\left( 
\begin{array}{cc} \Gamma^\chi_{11} & \Gamma^\chi_{12}\\ 
\Gamma^\chi_{21} & \Gamma^\chi_{22} \end{array} \right) 
\left( 
\begin{array}{c} J^0_1 \\ J^0_2 \end{array} \right)_R \chi^* +h.c.,
\nonumber\\
&& L^\rho_Y =(\bar l^0_1,\bar l^0_2,\bar l^0_3)_L\mu^\rho 
\left( 
\begin{array}{c} e^0_1 \\  e^0_2 \\ e^0_3 \end{array} \right)_R\rho 
+{\bar Q^0}_{3L}(\Gamma^\rho_{31},\Gamma^\rho_{32} \Gamma^\rho_{33})
\left( 
\begin{array}{c} d^0_1 \\ d^0_2 \\d^0_3 \end{array} \right)_R\rho \label
{4}\\ 
&&+(\bar Q^0_1,\bar Q^0_2)_L
\left( 
\begin{array}{ccc}
 \Gamma^\rho_{11} & \Gamma^\rho _{12} & \Gamma^\rho_{13}\\ 
\Gamma^\rho_{21} & \Gamma^\rho_{22} & \Gamma^\rho_{23} \end{array} 
\right) 
\left( \begin{array}{c}
 u^0_1 \\ u^0_2 \\ u^0_3  
 \end{array} \right)_R \rho^* +h.c.,\nonumber\\
&& L^\eta_Y =(\bar l^0_1,\bar l^0_2,\bar l^0_3)_L\mu^\eta 
\left( \begin{array}{c}
 \nu^0_1 \\  \nu^0_2 \\ \nu^0_3 \end{array} \right)_R\eta 
+{\bar Q^0}_{3L}(\Gamma^\eta_{31},\Gamma^\eta_{32} \Gamma^\eta_{33})
\left( \begin{array}{c}
 u^0_1 \\ u^0_2 \\ u^0_3 \end{array} \right)_R\eta \nonumber\\
&&+(\bar Q^0_1,\bar Q^0_2)_L
\left( \begin{array}{ccc}
 \Gamma^\eta_{11} & \Gamma^\eta _{12} & \Gamma^\eta_{13}\\ 
\Gamma^\eta_{21} & \Gamma^\eta_{22} & \Gamma^\eta_{23} 
\end{array} \right) 
\left( 
\begin{array}{c}
 d^0_1 \\ d^0_2 \\ d^0_3  
 \end{array} \right)_R \eta^* +h.c.,\nonumber\\
&& {\cal D}^\mu Q^0_{i} 
=(\partial^\mu +ig\frac {1}{2}\lambda^T\cdot A^\mu 
+i\frac {1}{3}g_NB^\mu)Q^0_{iL},\nonumber
\end{eqnarray}
where the expressions for the other covariant derivatives are omitted and 
$\mu^{(\chi,\rho,\eta )}$ denote $3\times 3 $ matrices. The Lagrangians 
except for the Yukawa interactions ${\cal L}_Y^{(\chi,\rho,\eta )}$, 
which are written in the case of the bosonic singlet $``1"$\cite{MIN,HL,RHN}, 
have the same form independent of an interpretation of the singlet. 
When the fermionic $(1_3)$ and antifermionic $(1_{\bar 3})$ singlets are adopted, 
the Yukawa interactions are obtained from above by putting 
$\Gamma^{( \rho, \eta )}_{3i}=\Gamma^{( \rho, \eta )}_{i3}=0 (i=1,2)$ 
because then the interactions such as 
$\sum_{i=1}^2 {\bar Q^0}_{3L}\Gamma_{3i}^{\rho}d^0_{iR}\rho $ and 
$\sum_{i=1}^2{\bar Q^0}_{iL}\Gamma_{i3}^{\rho}u^0_{3R}\rho^*$ are 
{\it not scalar} 
under a transformation of $\rm SU(3)_L$\cite{KKM}. It is noted that the expressions 
for the leptons are unique because only the fundamental representation for 
the left-handed representations is assigned and then the singlet is meant 
with respect to a transformation of the fundamental representation 
as stated above\cite{SU,PL}.

The Lagrangian $L_f$ in ({\ref 4}) for the basic fermions is rewritten 
in terms of the weak interaction bases ( with the suffices $0$ ) as follows
\begin{eqnarray}
&& L_f ={\rm kinetic\ energy\ terms}+eJ^{\mu}_{em}A_{\mu}\nonumber\\
&&+ \frac {g}{2\sqrt 2}
\left[J^{\mu}_WW^+_\mu +J^{\mu}_XX^+_\mu+J^{\mu}_YY^{++}_\mu +h.c.\right]
+\frac {g}{2c_W}J^{\mu}Z_\mu 
+\frac {g}{2\sqrt {1-3t^2_W}}J^{\mu}_{Z^\prime}Z_{\mu}^{\prime},\label
{5}
\end{eqnarray}
where
\begin{eqnarray*}
&& J^{\mu}_{em}=-{\bar e^0}_a\gamma^\mu e^0_a
+{\bar E^0}_a\gamma^\mu E^0_a 
+\frac {2}{3}{\bar u^0}_{a}\gamma^\mu u^0_a
-\frac {1}{3}{\bar d^0}_{a}\gamma^\mu d^0_a
+\frac {5}{3}{\bar J^0}_{3}\gamma^\mu J^0_3
-\frac {4}{3}{\bar J^0}_{i}\gamma^\mu J^0_i , \\  
&& J^{\mu}_W ={\bar l^0}_{aL}\gamma^\mu (\lambda_1+i\lambda_2)l^{0}_{aL}
+{\bar Q^0}_{3L}\gamma^\mu (\lambda_1+i\lambda_2)Q^{0}_{3L}
+{\bar Q^0}_{iL}\gamma^\mu (-\lambda^T_1-i\lambda^T_2)Q^{0}_{iL} ,\\  
&& J^{\mu}_X ={\bar l^0}_{aL}\gamma^\mu (\lambda_4-i\lambda_5)l^{0}_{aL}
+{\bar Q^0}_{3L}\gamma^\mu (\lambda_4-i\lambda_5)Q^{0}_{3L}
+{\bar Q^0}_{iL}\gamma^\mu (-\lambda^T_4+i\lambda^T_5)Q^{0}_{iL} ,\\
&& J^{\mu}_Y ={\bar l^0}_{aL}\gamma^\mu (\lambda_6-i\lambda_7)l^{0}_{aL}
+{\bar Q^0}_{3L}\gamma^\mu (\lambda_6-i\lambda_7)Q^{0}_{3L}
+{\bar Q^0}_{iL}\gamma^\mu (-\lambda^T_6+i\lambda^T_7)Q^{0}_{iL} ,\\
&& J^{\mu}_Z ={\bar l^0}_{aL}\gamma^\mu (\lambda_3)l^{0}_{aL}
+{\bar Q^0}_{3L}\gamma^\mu (\lambda_3)Q^{0}_{3L}
+{\bar Q^0}_{iL}\gamma^\mu (-\lambda^T_3)Q^{0}_{iL}-2s^2_WJ^\mu_{em} ,\\
&& J^{\mu}_{Z^\prime} ={\bar l^0}_{aL}\gamma^\mu (\lambda_8)l^{0}_{aL}
-{\sqrt 3}t^2_W{\bar l^0}_{aL}\gamma^\mu (\lambda_3)l^{0}_{aL}
+{\bar Q^0}_{3L}\gamma^\mu (\lambda_8)Q^{0}_{3L}
-{\sqrt 3}t^2_W{\bar Q^0}_{3L}\gamma^\mu (\lambda_3)Q^{0}_{3L}\\
&& \quad\quad +{\bar Q^0}_{iL}\gamma^\mu (-\lambda^T_8)Q^{0}_{iL}
-{\sqrt 3}t^2_W{\bar Q^0}_{iL}\gamma^\mu (-\lambda^T_3)Q^{0}_{iL}+2
{\sqrt 3}t^2_{W}J^\mu_{em} ,\\
&& e=\frac{gg_N}{\sqrt {g^2+4g^2_N}}=\frac{gg_Y}{\sqrt {g^2+g^2_Y}} , 
\qquad  
t_W\equiv \tan \theta_W=\frac {g_N}{\sqrt {g^2+3g^2_N}}.
\end{eqnarray*}
The gauge bosons can be expressed in terms of the original $A^i_\mu$'s,\ $B_\mu$ 
but their explicit forms are omitted here because it is easily known and 
trivial\cite{MIN,HL,RHN}. It is noted that as in the case of the standard model 
(SM)\cite{SM} all the right-handed fields are absorbed into the $J^{\mu}_{em}$ 
current together with the corresponding left-handed fields and only some left-handed 
fields remain in the expressions of the currents except for the $J^{\mu}_{em}$ current, 
and the current parts of the left-handed fields are given in the form sandwiching 
the representation or complex conjugate representation matrices between 
the two left-handed fields given in ({\ref 1}). 
The representation matrices of $3$ appear in the left-handed leptons and the quarks of 
the third family due to an assignment $3$ to these fields in ({\ref 1}), 
while the assignment of $\bar 3$ to the left-handed fields in the first 
and second families in ({\ref 1}) leads to an introduction of 
the complex conjugate representation matrices, i.e. transposed matrices 
with a minus sign. Thus, it will be necessary that the expressions for 
the current are given in the forms similar to the above ones in order for 
the anomaly coefficients even after the symmetry breaking to disappear, 
otherwise the anomaly coefficients can not be expressed in the form with 
the trace of the product of the representation matrices \cite{J}. It is, 
thus, expected that in order for the theory to be anomaly free and thus renormalizable 
the mass eigenstates (physical fields) after symmetry breaking be expressed 
in terms of a linear combination of the original fields in three families for the leptons, 
while in the case of the quarks those in the first and second families be given 
in terms of the original fields of the two families and those in the third family 
agree with the original fields corresponding to each. We will show 
that the anomaly coefficients for the possible fermion triangle diagrams 
can not be expressed in the form of the trace of only the product of 
the representation matrices in the case of the adoption of the bosonic singlet 
$``1"$ for the right-handed quarks as usually assigned\cite{MIN,HL,RHN} 
and in the case of $``1_3"$ and $``1_{\bar 3}"$ the anomalies do not appear producing 
the renormalizable results in a sense with no anomaly\cite{KKM}.

The masses for the leptons are obtained from the Yukawa interactions in 
({\ref 4}) in a similar way as in SM, and then the currents in ({\ref 5}) may be 
given by replacing the weak interaction bases $l^0_{aL} \rightarrow$ the bases 
$ l_{aL}$ corresponding to the mass eigenstates. Thus, the discussion on the leptons 
is the same as in SM and thus is omitted here \cite{MIN,HL,RHN}. 
Similarly, the quark masses are given from the Yukawa interactions in ({\ref 4}) 
by the vacuum expectation value (VEV) of $\chi,\ \rho$, and $\eta$.

The VEV of $\langle\chi\rangle = (0,0, \chi_v)^T/\sqrt 2$ gives the mass to 
the $J$'s quarks as follows
\begin{eqnarray}
m_{J_3}{\bar J}_3J_3 +{\bar J}M_JJ ,\label{6}
\end{eqnarray}
where $m_{J_3}(=\chi_v\Gamma^\chi_3/\sqrt 2)$ denotes the mass of 
$J_3 ( =J^0_3)$, and
\begin{eqnarray*}
&& J^0_{L,R}=A^J_{L,R}J_{L,R},
\quad {A^J}^\dag_L M^\chi A^J_R = M_J =\left(
\begin{array}{cc}
m_{J_1} & 0 \\
0 & m_{J_2}
\end{array}\right),\\
&& J^0_{L,R}=\left( 
\begin{array}{c} 
J^0_1 \\ 
J^0_2 
\end{array}\right)_{L,R}, \qquad M^\chi 
=\frac {1}{\sqrt 2}\chi_v\left( 
\begin{array}{cc} 
\Gamma^\chi_{11} & \Gamma^\chi_{12} \\ 
\Gamma^\chi_{21} & \Gamma^\chi_{22} 
\end{array} \right). 
\end{eqnarray*}
with the $2\times 2$ unitary matrices $A^J_{L,R}$. It is noted that 
the mass eigenstate of the $J_3$ quark in the third family is the same as 
the weak interaction basis $(J^0_3)$ but those ($J_1,J_2)$ in the first 
and second families are given by a linear combination of 
the weak interaction bases ( $J^0$'s ) in the families independently of 
an interpretation of the singlets. 
The result is due to the fact that the Yukawa interactions of the quarks with 
the $\chi $ are common in the cases of $``1"$ and $``1_{3}" (``1_{\bar 3}")$.

The VEV of $\langle\rho\rangle = (0, \rho_v,\ 0)^T/\sqrt 2$, together with 
that of $\langle\eta\rangle = (\eta_v,\ 0,\ 0)^T/\sqrt 2$, in ({\ref 4}) gives 
the masses to the $u$ and $d$ quarks in terms of the linear combinations of 
the quarks in the three families in order to forbid unphysical processes such as 
$t \longleftrightarrow (u\,, c)$,\ $b \longleftrightarrow (d\,, s)$ 
without interactions (or through interaction with vacuum) and then the 
mass terms of the $u$ and $d$ quarks are given from the Yukawa interactions 
in terms of a linear combination of the corresponding interaction bases 
in three families as follows
\begin{eqnarray}
{\bar U}_L M_uU_R +{\bar D}_L M_dD_R +h.c.  ,\label{7}
\end{eqnarray}
where
\begin{eqnarray*}
&& U^0_{L,R} = \left(
\begin{array}{c}
u^0_1 \\
u^0_2 \\
u^0_3
\end{array}\right)_{L,R},\quad\quad D^0_{L,R}=\left(
\begin{array}{c}
d^0_1 \\
d^0_2 \\
d^0_3
\end{array}\right)_{L,R},\\
&& U^0_{L,R}=B^u_{L,R}U_{L,R},\quad D^0_{L,R}
=B^d_{L,R}D_{L,R},\\
&& \frac{1}{\sqrt 2}{B^u}^\dag_L(I_2\Gamma^\rho\rho^0_v 
+ I_1\Gamma^\eta\eta^0_v)B^{u}_R =M_u = \left(
\begin{array}{ccc}
m_{u_1} & 0 & 0 \\
0 & m_{u_2} & 0 \\
0 & 0 & m_{u_3}
\end{array}\right),\\
&& \frac{1}{\sqrt 2}{B^d}^\dag_L
(I_1\Gamma^\rho\rho^0_v 
+ I_2\Gamma^\eta\eta^0_v)B^{d}_R = M_d 
= \left(
\begin{array}{ccc}
m_{d_1} & 0 & 0 \\
0 & m_{d_2} & 0 \\
0 & 0 & m_{d_3}
\end{array}\right),\\
&& B^{(u,d)\dag}_{L,R}B^{(u,d)}_{L,R} = \left(
\begin{array}{ccc}
1 & 0 & 0 \\
0 & 1 & 0 \\
0 & 0 & 1
\end{array}\right),\quad\quad  I_1=\left(
\begin{array}{ccc}
0 & 0 & 0 \\
0 & 0 & 0 \\
0 & 0 & 1
\end{array}\right),\quad\quad I_2=\left(
\begin{array}{ccc}
1 & 0 & 0 \\
0 & 1 & 0 \\
0 & 0 & 0
\end{array}\right).
\end{eqnarray*}
It is noted that the mass eigenstates, $U_{L,R}$ and $D_{L,R}$, of the $u$ and $d$ 
quarks are given in terms of a linear combination of the corresponding 
interaction bases in three families in contrast with the case of $J$'s and 
the following relations between the interaction bases and the mass eigenstate ones hold
\begin{eqnarray}
U^{0\dag}_{L,R}U^0_{L,R} 
=U^\dag_{L,R}U_{L,R} {\rm and} 
D^{0\dag}_{L,R}D^0_{L,R} 
=D^\dag_{L,R}D_{L,R}.\label{8}
\end{eqnarray}

As will be seen, (8) means that the anomaly coefficients after the symmetry breaking 
can not be expressed in terms of product of the representation matrices.

In the case of an adoption of $1_3$ and $1_{\bar 3}$ for the singlets, 
it follows that the mass eigenstates for the $u$ and $d$ quarks are given 
as in the case of $J$'s because the mass matrices are given explicitly 
by putting $\Gamma^{\rho,\eta }_{3i}
=\Gamma^{\rho,\eta }_{i3}=0\ (i=1,2)$ 
in the above expressions as follows
\begin{eqnarray*}
&& \rho_vI_2\Gamma^\rho + \eta_vI_1\Gamma^\eta = \rho_v 
\left ( \begin{array}{ccc} 
\Gamma^\rho_{11} & \Gamma^\rho_{12} &  0 \\
\Gamma^\rho_{21} & \Gamma^\rho_{22} & 0 \\
0 &0 & 0 \end{array} \right) 
+ \eta_v \left( \begin{array}{ccc}
0 & 0 & 0 \\
0  & 0  & 0 \\ 
0 &0 & \Gamma^\eta_{33} \end{array} \right) ,\\   
&& \rho_vI_1\Gamma^\rho +\eta_vI_2\Gamma^\eta =
 (\ \rho \ \Longleftrightarrow\ \eta\ {\rm  in\ above }\ ).
\end{eqnarray*}
It is evident from the expressions that the relations corresponding to ({\ref 8}) 
for the mass eigenstates split into two parts, one for a linear combination of 
the interaction bases in the first and second families and only the other one 
in the third family as in the case of $J$'s. Explicitly, the followings hold
\begin{eqnarray}
&& u^\dag_L m_u u_L +u^\dag_{3L} m_{u_3}u_{3R}+  d^\dag_L m_d d_L 
+d^\dag_{3L}m_{d_3}d_{3R}+h.c.,\nonumber\\
&& u^0_{L,R}=A^u_{L,R}u_{L,R}, \qquad d^0_{L,R}=A^d_{L,R}d_{L,R},
\nonumber\\
&& \frac {1}{\sqrt 2}\rho_vA^{u\dag}_L\left(
 \begin{array}{cc}
 \Gamma^\rho_{11} & \Gamma^\rho_{12} \\
 \Gamma^\rho_{21} & \Gamma^\rho_{22}
 \end{array} \right) A^u_R = m_u =diagonal,\qquad u_{3} =u^0_{3},
\nonumber\\
&& \frac {1}{\sqrt 2}\eta_vA^{d\dag}_L\left(
 \begin{array}{cc}
 \Gamma^\eta_{11} & \Gamma^\eta_{12} \\
 \Gamma^\eta_{21} & \Gamma^\eta_{22}
 \end{array} \right) A^d_R 
 = m_d =diagonal,\qquad d_{3}=d^0_{3},\nonumber\\
&& u^{0\dag}_{L,R}u^0_{L,R} 
=u^{\dag}_{L,R}u_{L,R},\qquad d^{0\dag}_{L,R}d^0_{L,R} 
=d^{\dag}_{L,R}d_{L,R},\label{9}\\
&& u^{0\dag}_{3L,R}u^0_{3L,R}
=u^{\dag}_{3L,R}u_{3L,R},\qquad d^{0\dag}_{3L,R}d^0_{3L,R}
=d^{\dag}_{3L,R}d_{3L,R},\nonumber\\
&& u^0_{L,R} =\left(
 \begin{array}{c} u^0_{1} \\ u^0_{2}
 \end{array} \right)_{L,R},\qquad  d^0_{L,R} =\left(
 \begin{array}{c} d^0_{1} \\ d^0_{2}
 \end{array} \right)_{L,R}.\nonumber
\end{eqnarray}
The result is desirable from a physical point of view as mentioned before 
and is necessary for the anomaly coefficients after the symmetry breaking to 
have expressions corresponding to those in terms of the representations matrices 
after the symmetry breaking.

The currents in ({\ref 5}) in terms of the above mass eigenstates in ({\ref 7}) 
are expressed by replacing the weak interaction bases $l^0_{aL}$
$\rightarrow$ the bases containing the lepton mass eigenstate $l_{aL}$ 
for the leptons, while the expressions for the quark parts in 
the $J^\mu_W,\ J^\mu_X,\ J^\mu_Y,\ J^\mu_Z,\ J^\mu_{Z^\prime}$, except for 
the current $J^\mu_{em}$ which is diagonal in the quark flavor and thus is given by 
the same form as the original one due to ({\ref 8}), can not be obtained only by 
replacing the weak interaction bases $Q^0_{aL}\rightarrow$ the mass eigenstate bases 
consisting of $Q_{aL}$ because the quarks in the first and second families are mixed with 
the corresponding one in the third family and the relations ({\ref 8}) must hold in 
contrast with that in the case of $J$'s. 
That is, in the case of  the bosonic singlet ($``1"$) the expressions for 
the quark currents (except for $J^\mu_{em}$) after the spontaneous symmetry breaking 
can not be expressed in the forms similar to those before the symmetry breaking 
though the lepton parts have the similar form in all currents due to the assignment of 
only the fundamental representation to three families. 
Thus, it follows that the anomaly coefficients can not be expressed in terms of 
the trace of the products of the representation matrices for the possible triangle diagrams 
and then the anomalies will appear bringing about non-renormalizability of the theory 
in a sense with anomaly. Furthermore, it is noted that the flavor changing neutral current 
(FCNC)\cite{FCNC} as well as CP(T) violation\cite{CP} through the lepton 
and quark parts appears in this case. It is known that FCNC can be avoided by 
taking into account the horizontal symmetry in addition to the 331 model\cite{ho}.

On the other hand, in the case of the adoption of the fermionic ($``1_3"$) and 
antifermionic ($``1_{\bar 3}")$ singlets, all currents for the quark parts in terms of 
the mass eigenstates are given in the similar form only with replacements by 
the weak interaction bases $\rightarrow $ the bases containing the mass 
eigenstates as in SM and thus it is obvious from the relations ({\ref 9}) in this case 
that the triangle anomalies for all possible diagrams do not appear from vertex of 
three currents as easily seen from the anomaly coefficients given by the trace of 
the product of the representation matrices for the fermion fields. 
The currents in terms of the mass eigenstates are given only by replacing 
the weak interaction bases with the bases in terms of the mass engenstate one's in 
(\ref{9}) as in the case of SM and some of them are as follows
\begin{eqnarray*}
&& J^{\mu}_W ={\bar l}_{aL}\gamma^\mu (\lambda_1+i\lambda_2)l_{aL}
+{\bar Q}_{3L}\gamma^\mu (\lambda_1+i\lambda_2)Q_{3L}
+{\bar Q}_{iL}\gamma^\mu (-\lambda^T_1-i\lambda^T_2)Q_{iL} ,\\ 
&& J^{\mu}_Y ={\bar l}_{aL}\gamma^\mu (\lambda_6-i\lambda_7)l_{aL}
+{\bar Q}_{3L}\gamma^\mu (\lambda_6-i\lambda_7)Q_{3L}
+{\bar Q}_{iL}\gamma^\mu (-\lambda^T_6+i\lambda^T_7)Q_{iL} ,\\
&& J^{\mu}_Z ={\bar l}_{aL}\gamma^\mu (\lambda_3)l_{aL}
+{\bar Q}_{3L}\gamma^\mu (\lambda_3)Q_{3L}
+{\bar Q}_{iL}\gamma^\mu (-\lambda^T_3)Q_{iL}-2s^2_WJ^\mu_{em} ,
\end{eqnarray*}
where
\begin{eqnarray*}
&& l_{aL}=\left( 
\begin{array}{c} \nu_a \\ e^\prime_a \\  E^\prime_a 
\end{array} \right)_L ,
Q_{3L}=\left( 
\begin{array}{c}
 u_3 \\ d_3 \\  J_3 \end{array} \right)_L , 
Q_{iL}=\left( 
\begin{array}{c}
 d_i \\ u^\prime_i \\  J^\prime_i \end{array} \right)_L , \\
&& e^\prime_L = U_{\nu e}e_L, E^\prime_L 
= U_{\nu E}E_L,
u^\prime_L = U_{d u}u_L,
J^\prime_L = U_{d J}J_L.
\end{eqnarray*}
The other currents as well as the $J^\mu_{em}$ current are expressed in 
the same way by these bases. 
It is noted that the $3\times 3$ unitary matrices $U_{\nu e}$ and $ U_{\nu E}$ are 
the Cabibbo-Kobayashi-Maskawa matrices\cite{CKM} for the leptons 
and the $2\times 2$ unitary matrices $ U_{d u}$ and $U_{d J}$ denote 
the Cabibbo matrices\cite{CM}. The $J^{\mu}_{em}$ current may be given in terms of 
the above bases and the right-handed one or only in terms of the mass eigenstate bases 
instead of the weak interaction bases because the $J^\mu_{em}$ is diagonal in the flavor. 
Thus, it is evident that the anomaly coefficients are given by the same form as 
in the case of the weak interaction bases due to the following relations 
\begin{eqnarray*}
&& e^{\prime\dag}_L e^{\prime}_L =e^\dag_Le_L=e^{0\dag}_L e^0_L,
E^{\prime\dag}_LE^{\prime}_L=E^{\dag}_LE_L=E^{0\dag}_LE^0_L,
\nu^{\dag}_L\nu_L=\nu^{0\dag}_L\nu^0_L\\
&& u^{\prime\dag}_Lu^{\prime}_L=u^{\dag}_Lu_L=u^{0\dag}_Lu^0_L,
J^{\prime\dag}_LJ^{\prime}_L=J^{\dag}_LJ_L=J^{0\dag}_LJ^0_L,
d^{\dag}_Ld_L=d^{0\dag}_Ld^0_L,
\end{eqnarray*} 
and the mass eigenstates of the quarks in the third family with the same bases as 
in the original weak interaction bases. The mass eigenstates corresponding to 
the neutral gauge bosons $Z_\mu$ and $Z^\prime_\mu$ are given in terms of 
a linear combination of $Z_\mu$ and $Z^\prime_\mu$ and thus the anomaly coefficients 
for the related processes in terms of the mass eigenstates become zero 
if those in the case of $Z_\mu$ and $Z^\prime_\mu$ is zero. 
It is noted that the anomaly coefficients are not necessarily expressed in terms of 
the sum over the whole charges in three families in contrast with those in SM. 
For instance, for the process $Z^\prime \rightarrow Y^{++}Y^{--}$ 
the anomaly coefficient becomes essentially
\begin{eqnarray*}
\sum {\rm tr} T_{8L}
\{T_{6L}-iT_{7L},T_{6L}+iT_{7L}\}
&=& \frac {3}{8\sqrt 3}
\left[ {\rm tr} \lambda_8
\{\lambda_6+i\lambda_7,\lambda_6-i\lambda_7\}\right. \\
&&\left. +{\rm tr} \lambda_8
\{\lambda_6+i\lambda_7,\lambda_6-i\lambda_7\}\right. \\
&&\left. +2{\rm tr} (-\lambda^T_8)
\{-\lambda^T_6-i\lambda^T_7,-\lambda^T_6+i\lambda^T_7\}\right] ,
\end{eqnarray*}
 where the first term expresses a contribution from the 3 families of the leptons 
and the representation matrices $\lambda_i$ $( i=6, 7, 8 )$, 
the second term from the third family of the quark with 3 colors 
and the representation matrices $\lambda_i$,  and the third term from 
the first and second families of the quark with each 3 colors 
and the representation matrices $-\lambda^T_i$. 
It, thus, follows that the result becomes zero due to the sum over not 
all charge on the three families.  The FCNC does not appear here and CP(T) 
violation appears through the lepton parts only.

It may be concluded that in order for the 331 like-model to be 
renormalizable even after the spontaneous symmetry breaking as well as 
before that the Yukawa interactions must be given in the form with
 $\Gamma_{3i}=\Gamma_{i3}=0\ (i=1,2)$ in ({\ref 4}) which requires 
the assignment of the singlet to the right-handed quarks accompanied 
with the left-handed ones should be distinguished whether 
the right-handed singlet (scalar) is a counterpart of the left-handed 
$3$ or $\bar 3$. In this case, the singlet accompanied with 
the left-handed $3$ is only $``1_3"$ and that with $\bar 3$ is $``1_{\bar 3}"$, 
i.e. a scalar under a transformation of the fundamental representation $3$ 
but not under ${\bar 3}$ for $``1_3"$ and vice versa for $``1_{\bar 3}"$. 
Then, the renormalizability of the theory will be guaranteed in a sense of 
no anomalies even after the gauge fixing and of course FCNC as well as 
the triangle anomalies does not appear  in contrast with use of only 
the singlet $``1"$ for the right-handed counterparts of the left-handed 
$3$ and ${\bar 3}$. 

The above is discussed without fixing the gauge and detailed analysis 
with $R_\xi$ gauges will be given somewhere because only papers based on 
Higgs mechanism \cite{MIN,HL,RHN} seems to be existing as far as we know.

\vspace{4mm}
\noindent
{\bf Acknowledgments} :
One of the authors (T.K.) is grateful to Prof T. Maekawa for valuable comments, 
helpful discussions and continuous encouragement.

\end{document}